\documentclass[12pt,preprint]{aastex}

\shorttitle{Extra galactic high redshift objects}
\shortauthors{Navia et al.}

\begin{document}

\title{The Hubble diagram of high redshift objects, QSOs and AGNs}


\author{C. E. Navia, C. R. A. Augusto, and K. H. Tsui}
\affil{Departamento de Fisica, Universidade Federal Fluminense,
    Niteroi, RJ, Brazil 24210-130}

\begin{abstract}
According to the Hubble law, high redshift objects such as Quasar (QSOs), X-ray Active Galactic Nuclei (AGN) together with the Gamma Ray Burst (GRBs) are the fastest and farthest objects. These characteristics provides strong motivations for they to be used to constrain the cosmological parameters, without the limitations found in the Ia supernovae study and which is restricted to redshift of up to 1.7. 
However, the variability and behavior in the QSOs and AGNs spectra tell us that they have very complex structures and the standard candle framework can not be applied.  So far the available data of QSOs and AGNs have shown some anomalies observed in their  brightness and metallicities, difficult to understand, under an orthodox point of view. Here, we show that their Hubble diagram flattens for $z\geq 3 $. The result need further confirmation, because the statistics of high redshift extragalactic objects is still poor. Details and some implications of these results are reported in this work.
\end{abstract}


\section{Introduction}

The quasars (QSOs) were discovered by radio telescopes in the late 1950s. Many were recorded as radio sources with no corresponding visible object. It is believed that the QSOs are
extremely bright and distant active galactic nuclei (AGN) of young galaxies. They were first identified as being high-redshift sources of electromagnetic energy, including radio waves and visible light that were point-like. More than 60,000 quasars are known, all observed spectra have shown considerable redshifts, ranging from 0.06 to the recent maximum of 6.4. Quasars are found to vary in luminosity on a variety of time scales. The Sloan Digital Survey has uncovered the most distant quasars known (up to $z\sim 6.4$). In addition, there is also the report of high redshift X-ray AGNs with moderate to high quality spectra observed by XMM-Newton and Chandra.

On the other hand, a study made by Gehren et al \citep{gehren84} has shown that
quasars are usually located in groups or clusters of galaxies which are dominated by the luminosity contribution of the quasar host galaxy. It is suggested that most radio-loud quasars are situated in giant elliptical galaxies in which a considerable amount of hot gas and young stars may be present owing to recent collisions or tidal interaction with nearby cluster members.
In addition, from a study of pairs of quasars and galaxies, which were very close in the sky, \citep{stockton78} 
found that, in a fraction ($\sim 50\%$) of the sample of quasar-galaxy, the redshifts of the galaxies agreed with the redshifts of the quasars. In other words, quasars are associated with galaxies that have the same redshift as the quasar, and have just the brightness expected, if the quasars are at their cosmological distances. However, according to the standard cosmology, the study
can include only quasars at low or moderate redshifts, because the $(1+z)^{-4}$ dimming in surface brightness makes normal galaxies unobservable at large redshifts \citep{lehnert92}. 

However, if the quasar-galaxy pair has different redshifts ($z_q>z_G$), just as it happens in $\sim 50\%$ of the Stockton's sample, the situation becomes controversial. Burbidge and coworkers have shown \citep{burbidge90} that there is strong evidence that normal galaxies and QSOs tend to clustered, whether or not their redshifts are the same. While
if the quasar redshift is a distance indicator, as is advocated by the standard cosmology,
the identification of quasars shining through or close to galaxies offer a way to probe gravitational lensing effects. 
From quasar-galaxy gravitational lensing, is possible to deduce that in some quasar-galaxy pairs with very different redshifts, the association is only apparent. Despite the small angular separation between the quasar and its galaxy companion, the quasar must be beyond the lens galaxy. The method works only for galaxy lenses, with $z\geq 0.5$ \citep{dar91}. However, a more accurate analysis on gravitational lensing effect shows that gravitational lens measurements can be a powerful tool, but only if a variety of major difficulties are overcome, such as the 
multiple lenses and microlensing background and the small variability amplitude typical of most quasar \citep{shechter04}.

In this paper, we have made an analysis on the Hubble diagram of extragalactic high-z objects (QSOs and AGNs), on the basis of the more recent data as possible. 
This paper is organized as follows: In Sec.2, extra galactic objects such as QSOs and AGNs at high redshift are described. In Sec.3, the extragalactic objects in a cosmological context is presented, ana Sec.4, contains our conclusions.

\section{QSOs and AGNs features at high redshift}

In spite of some progresses obtained on the nature of the quasars, some subjects of them still remain open,
such as: a qualitative test of quasar evolution through their spectra \citep{segal98} have shown that spectra of low and high  redshift quasars are very similar. In the first order, there is nothing that distinguishes high redshift quasar from low redshift quasars. In addition, the observation of quasar APM 8279+5255 with a redshift of $z=3.51$ by ESA's XMM-Newton satellite \citep{hasinger02}, had shown 
that iron was three times more abundant in the quasar than in our Solar System. If the redshift is indicative of distance, as an object formed to $13.5\times 10^9$ years ago (90 \% of the age of the universe) can it contain 3 times more iron that the solar system only formed to only $5\times 10^9$ years ago?. 
 Similar results has been found in recent studies based on quasar emission lines, the quasar environments are typically metal rich, with metallicities near or above the solar value at even the highest observed redshifts \citep{simon07} 

In the standard cosmology, the high redshift of the quasars is an indicator that they are between the most distant objects in the universe.  They must, therefore, emit huge amounts of energy in order to be visible at such colossal vast distances, there is not a known mechanism to generate these colossal energy sources, because the thermonuclear source would be insufficient. In addition, due to their variability in luminosity on a variety of time scales, they must be relative small objects. A huge amount of matter in a small space might look like a black hole. Thus to explain the quasar source energy responsible of its high observed luminosity require to postulate that quasar and AGNs are powered by gravity. Super-massive black holes devouring large amounts of material organized in a thin accretion disk \citep{elvis00}. Even so, the profiles of emission line in the spectra of most quasars and AGNs are centrally peaked, telling us that only minor fractions of this light come from material at very high velocities. This sounds more like an amorphous cloud, or maybe a spherical distribution of clouds, than a thin, organized disk. Others problems arise with the super-massive black hole hypothesis, because a well accepted property of black holes is that 
anything that gets within a certain distance of the black hole's center, called the event horizon, will be trapped. Consequently,
they cannot sustain a magnetic field of their own. But observations of quasar Q0957+561 \citep{schild05} situated $7.8$ billion light years indicate that the object powering it does have a magnetic field. Those details suggest that the central object is not a black hole, for this reason, has been suggested that rather than a black hole, this quasar contains something called a magnetospheric eternally collapsing object (MECO) \citep{robertson03}.

In addition, if the quasar redshift is a distance indicator, this implies that objects like quasars with high redshifts must be superluminal objects, because the cosmological distance per unit cosmic time (the so called recession velocity) in the FRW metric is given by
\citep{davis03}
\begin{equation}
v_{rec}(t,z)=\frac{c}{R_0}\dot{R} (t)\int_0^z\frac{dz'}{H(z')},
\end{equation}
where the  expansion history of the universe is given by $H(z)$. For a flat universe, there is the constrain condition for the cosmological density parameters as$\Omega_M+\Omega_{\Lambda}=1$ and $H(z)$ is given by
\begin{equation}
H(z)^2=H_0^2\left[(1+z)^3 \Omega_M+\Omega_{\Lambda}(1+z)^{3(1+w)}\right],
\end{equation}
where $H_0$ is the Hubble constant \citep{olive06}. Fig.1 shows the velocity recession as a function of the redshift according to
the standard cosmology. We have used the  canonical cosmological parameters, close to the found in the CMB WMAP data,
$\Omega_M=0.3$, $\Omega_{\Lambda}=0.7$ and $w=-1$ for the dark energy parameter. Fig.1 shows also for comparison, the Special Relativity Theory (SRT) prediction
and the $cz$ approximation for the recession velocity. From this picture is possible to see that in the standard cosmology the velocity of recession is greater than c for $z \geq 1.407$. It is claim that this results is not a contradiction of special relativity theory, because the SRT is valid only under a flat Minkoski's space-time metric. Even so, the apparent magnitude of high redshift objects  like supernovaes or quasars, rule out the SRT Doppler interpretation of cosmological redshift at a confidence level of $23\sigma$
\citep{davis03}.

In astrophysics superluminal motion is seen as an ejection of mass (plasma) forming jets in some galaxies, AGNs and QSOs. Rees has shown  \citep{rees66} that an object moving relativistically in suitable directions may appear to a very distant observer to have a transverse velocity much greater than the velocity of light, consequently the superluminal motion will be only apparent. However, the superluminal motion has been seen also in a local nearby (galactic) source \citep{mirabel94}. On the other hand,
in order to explain the origin of the highest cosmic rays and neutrino physics, has bean suggesting a possible Lorentz symmetry violation \citep{coleman99}. This means that at high energies the speed of a proton were higher (bigger than c). Then it turns out that the proton would lose energy to electromagnetic radiation, until its speed was the same as the speed of light. Thus, the
Lorentz symmetry can not be a general symmetry of nature. 

On the other hand, the Gunn-Peterson trough in the quasar absorption spectra is recognised as a signature of intergalactic neutral hydrogen \citep{gunn65}. While,
no such effect was visible in the spectrum of quasar with $z<6$, implying that intergalactic hydrogen is practically all ionized up to a very high redshift. However, from about 2001, SDSS began to turn up quasars with $z \geq 6$ \citep{fan06}, which did indeed show a Gunn-Peterson trough in their spectra. Thus, a redshift of about 6 seems to relate to the re-ionization epoch. While, analysis of the first year WMAP data on the polarization of the CMB  \citep{kogut03} implies that, if reionization was a simple single-step process, it must have been completed at a much higher redshift, $z = 17\pm 3$, around $\sim 65\%$ higher than the value obtained with the quasar data ($z\sim 6$). 
Furthermore, a study of galaxies in the redshift range $5.7-6.5$ \citep{malhotra04} has shown a value close to quasar data. According to this survey, the re-ionization was practically complete at $z \sim 6.5$. 
Recently, four quasar above redshift 6 were discovered by the Canada-French High-z Quasar Survey (CFHQS)\citep{chriss07}. The  mean millimeter continuum flux for CFHQS quasar is substantially lower than that for SDSS quasar at the same redshift and these new data suggest a reonization era at $z>6$.

\section{Extragalactic objects in a cosmological context}  

The publication of Hubble's article in 1929 \citep{hubble29} is the birth of modern observational cosmology.
Hubble compared his distances to Slipher's measurements of redshift and made a famous plot, which today is called the Hubble diagram. Hubble diagram shows that a galaxy's redshift increases linearly with its distance from Earth. 
Under the assumption that the galaxy's redshift is due to the Doppler effect, the farther away a galaxy is, the faster it moves away from us, this tells us that the Universe is in expansion.

Extensions of Hubble's work with spacecraft instruments, such as the Hubble Space telescope (Key Project) \citep{freedman01}, gives for the Hubble constant the value of $h = 0.71 \pm 0.02 (random) \pm 0.06(systematic)$. This large value for the Hubble constant when compared with its previous accepted value, suggests that beyond the nearby linear expansion of Hubble's law, the Universe expansion is accelerated.
This acceleration suggests that the other $\sim 70\%$ of the universe is composed of a ``dark energy'' that must have a negative pressure to make cosmic expansion speed up over time. 

However, a recent analysis also on the basis of the Hubble Space Telescope 
\citep{sandage06} has shown a discordant value to the Hubble's constant as $h\sim 62$, 12.7\% smaller than the value obtained by Key Project. We argue here that these two differing values can be correct if we assume a fractal structure for the cluster of galaxies, because the assumed fractal structure is now well supported by observations. Looking into different directions of the sky, and using the same technique, is possible to obtain different values for the Hubble constant. The discrepancy increases, taking into account different methods to measure distances.    
Measures of the Hubble constant, in the last decades of last century are framed in a strip of $h\sim 50$ to $h\sim 90$. The origin of these discordant values is probably due to the cluster of galaxies look like an infinitely replicated fractal structure.

 Is it possible to conciliate the fractal structure with the homogeneity suggested by the linear Hubble law? So far this issue is still a matter of considerable debate, such as the presence of an eventual cross-over to homogeneity. It has been suggested \citep{joyce00} that the fractal behavior is only a perturbation  inside of a homogeneous universe, in which the leading homogeneous component is the cosmic microwave background radiation (CMB). However, the situation becomes controversial with the recent results on
 the anisotropy of the CMB, such as the North-Souther asymmetry, with a Southern excess (in galactic coordinates) 
 \citep{tegmark03,wibig05}, sometimes attributed  only for the foreground. There is also gigantic voids in the CMB map \citep{rudnick07} identified also in radio waves, the biggest void exceeds by far the size of known region of empty space and also the expectations of computer simulations. The so called axis of evil in the CMB, initially observed in the COBE data as an alignment
 between the quadrupole and octopole \citep{lerner95} and also found in the WMAP data \citep{oliveira04,schwarz04,land05}.
This means that the universe is in fact arrayed around a special axis. If true, this behavior of the CBM constitute a challenge to the Cosmological principle.

 On the other hand, the SNIa supernovae are objects that appear when the mass of an accreting white dwarf increases to the Chandrasekhar limit and explode. The tight correlation of the luminosity indicators, such as the $L_p \sim \Delta m_{15}$ 
\citep{phillips93} and the multi color light curve shape \citep{riess95}, justify the use of SNIa
as standard candles and they are used to constrain the cosmological parameters \citep{perlmutter04,riess04}. The disadvantage in using the SNIa for cosmological purposes is their limited redshift, up to $z\sim 1.7$. Most of them are situated below $z<1$, At high redshift they are unobservable, due to the dust extinction in the intergalactic space. 
Therefore, in order to get more information in the high redshift region ($z>1$), it is necessary to look for other brighter astronomic objects such as GRBs \citep{hurley95}, Quasar and AGNs \citep{stocke92}.

In most cases, the Hubble diagram using quasars as cosmological objects, is presented only as a correlation between redshift and the apparent magnitude \citep{hewitt93,basu03}, because there is no direct information on quasars absolute magnitude. Even so, the Hubble plot of redshift against apparent magnitude is characterized by a large scatter. 
The absolute magnitude, called also as the quasar luminosity function, is derived from two observational parameters, the redshift (z) and the apparent magnitude (m), plus a cosmological model to obtain the so called 
luminosity distance ($d_L$). Thus in most cases the quasar luminosity function is given by
\begin{equation}
M=m+5-5 \log(d_L)-k+\Delta m(z),
\end{equation}
where $k=-2.5 \log (1+z)^{1-\alpha}$, $\Delta m(z)$ is the correction to $k$ taking into account the fact that the spectrum of quasar is not strictly a power law as the form $S \propto \nu^{-\alpha}$ with $\alpha=0.5$. Most of the studies of quasars, the luminosity distance $d_L$ is determined using  the Einstein de Sitter cosmological model where $\Omega_M=1$ and $H_0=50\;km\;s^{-1}Mpc^{-1}$, hereafter called as $\Omega$ model. However, there is also the $\Lambda$ dominant flat cosmology \citep{ostriker95} with $\Omega_M=0.35$, $\Omega_{\Lambda}=0.65$ and $H_0=65\;km\;s^{-1}Mpc^{-1}$, hereafter called as $\Lambda$ model, and it is included also the so called standard cosmology model with  
$\Omega_M=0.3$, $\Omega_{\Lambda}=0.7$ and $H_0=70km\;s^{-1}Mpc^{-1}$.

The present analysis on the quasar Hubble's diagram is made using the following data:
(a) The First Bright Quasar Survey (FBQS) catalog \citep{white00} has a set of 636 quasars distributed over 2682 $deg^2$.
An efficient selection criteria on the basis of artificial intelligence methods has been used to select the fraction of FBQS candidates that turn out to be quasars. The accuracy of this classification, has been tested using fivefold cross-validation. Most of the selected objects in the FBQS sample have 80\% of probability to be true quasars. Even so,
the FBQS classified any object with broad emission lines as a quasar, there are no distinction between quasars and
Seyfert 1 galaxies and quasars and broad lines radio galaxies. The absolute blue magnitude is given for objects with redshifts, and only a conventional cut at $M_B=-23$ was made. Of the 636 FBQS quasars, 50 so fall into this lower luminosity objects. The redshift for quasars and BL Lac objects in the FBQS sample were computed by cross-correlating the spectra with templates. The FBQS catalog includes radio-loud and radio-quiet quasars, the samples include also $\sim 29$ broad absorption line quasars and a number of new objects with remarkable optical spectra. In order to improve the uniformity of the samples, the quasars in the FBQS catalog were selected according to four criteria, including an extinction correction in the computed magnitude due to the Galactic latitude.

(b) The Fifth Data Release of the SLOAN DIGITAL SURVEY (SDSS)\citep{adelman07} . It has more than 65,450 objects with luminosities larger than $M_i=-22$ (in a standard cosmology) \citep{schneider95}. The catalog covers an area of $8000 deg^2$ and includes active galactic nuclei such as type II quasar, Seyfert galaxies and BL Lacertae objects.

(c) The 9th Veron-Cety$\&$Veron catalog, available in $www.obs-hp.fr$, includes 13214 QSOs, 462 BL Lac objects and 4428 AGNs (of which 1771 are Seyfert 1). These objects have broad emission lines, brighter than absolute magnitude $M_B=-23$ under a model close to $\Omega$ model.

In Fig.2 we show the apparent magnitude, the so called PSF magnitude obtained in the SDSS Release 5 photometric measurements \citep{adelman07} against the redshift (top panel in Fig.2). The figure includes (bottom panel Fig.2) the absolute magnitude as obtained by the SDSS data 
plus the standard cosmology model ($\Omega_M=0.3$, $\Omega_{\Lambda}=0.7$ and $H_0=70km\;s^{-1}Mpc^{-1}$). As is well known, these correlations show very large scatter. It is possible to observer (top panel) in the SDSS data a ``duck beak'' in the correlation between redshift and the apparent magnitude of quasars
at $z\sim 3$ as a transition for a flat correlation. Or in other words,
the apparent magnitude is consistent (statistically) with an almost constant value for $z>3$, independent of the value of the redshift.
  
In an attempt to improve the photometric accuracy and uniformity of the FBQS sample, the apparent magnitudes were recalibrated plate-by-plate using magnitudes from the Minnnesota automated Plate Scanner $POSS-I$ using a new limit of 17.8 in magnitude to redefine the complete sample. In addition, to improve the uniformity of the sample with Galactic latitude, an extinction correction was computed for each candidate object using the map of Schlegel  \citep{schlegel98}. In Fig.3 (top panel) it is possible to observe that the scatter of the absolute magnitude vs the redshift in the FBQS sample is less than that in the VC$\&$V sample shown in Fig.3 (bottom panel). In both samples, the $\Omega$ model is used to obtain the absolute magnitude. A similar  upside ``duck beak'' at $z\sim 3$ is also observed especially in the   $Veron\& Veron$ data as a transition for a flat correlation. This behavior is more clear in the lower picture because the VC$\&$V sample  extends to values of redshift bigger than in the FBQS sample. 

In order to extend the high quality FBQS sample for bigger values of redshift, other data taken from the literature were included, such as
the biggest redshift quasar ($z>5.8$) of the SDSS catalog \citep{fan01}. 
In these quasars, the absolute magnitude has been derived from a selection function $p(M_{1450},z)$, where $M_{1450}$ is the apparent magnitude in the rest frame at $1450 \AA$, and is calculated using a Monte Carlo Simulation of quasar color, based on the quasar spectral model and the two cosmological models $\Omega$ and $\Lambda$ respectively. We have also included six new quasars ($z\sim 6$) selected from $260 deg^2$ SDSS Southern survey,
a deep imaging survey obtained  by repeatedly scanning along the celestial Equator \citep{jiang07}, where the absolute magnitude or luminosity quasar function is derived from quasar selection function of $M_{1450}$ and $z$ in the standard cosmological model.

The analysis includes also moderate to high quality X-ray spectra of 10 of the most luminous AGNs,
at $z>4$. Nine are from XMM-Newton observation and one is from Chandra \citep{shemmer05}, where the apparent magnitude $AB_{1450}$ is measured and the absolute magnitude is derived on the basis of the standard cosmology with
$\Omega_M=0.3$, $\Omega_{\Lambda}=0.7$ and $H_0=70km\;s^{-1}Mpc^{-1}$.

Fig.4  shows the results of the apparent magnitude (central panel) and absolute magnitude (bottom panel) for quasars and AGNs respectively. The open red-square data for $z<4$ correspond to the quasars from FBQS catalog \citep{white00}, and the absolute magnitude is derived using the $\Omega$ cosmological model. The data for $z>5$ correspond to the quasars from SDSS catalog \citep{fan01}, and the absolute magnitude is derived using the $\Omega$ and $\Lambda$ model respectively. Finally, 
the solid red-square represent the X-ray AGN data, and the solid black-down triangles represent the SDSS Southern data respectively. In both cases, the absolute magnitude is obtained on the basis of the standard cosmological model. 
 The top panel in Fig.4 represents the magnitude modulus obtained as $m-M$ and is close to the logarithm of the luminosity distance, the solid-black line and the dashed-red line represent the prediction
of the standard model for two sets of cosmological parameters.

\section{Conclusions}

The high redshift objects such as QSOs, AGNs and GRBs open the doors for the study of the universe in another scale, the so called ``high-z universe''. So far, there are many controversial aspects on the nature of these high redshift objects.
The difficulties lie in the fact that for instance the QSOs and AGNs searches are still quite inhomogeneous over the sky, and
probably only large-scale automated surveys will be able to resolve this, and to answer some questions such as:
Is there evidence of quasars occurring in the direction of nearby galaxies? Can such an excess be explained by something like gravitational lensing? 

Recently, the concept of standard candle was applied to the GRB studies \citep{shaefer03,shaefer07,ghirlanda04,ghirlanda06}. In this approach the Hubble diagram and the cosmological 
parameters were obtained. It's claimed an agreement with the standard cosmology inside  $2\sigma$ to $3\sigma$ confidence level. 
The information that is obtained from GRBs is still dependent on the concept of standard candle. Even so, remarkable effort has been made to calibrate the GRBs and to place them in conditions to constrain the cosmological parameters. In addition, the statistic is still incipient in the high redshift region to reach a more robust conclusion.

So far, QSOs, AGNs and GRBs have been observed with redshifts beyond 6. 
Unhappily up to now, the QSO and AGN luminosity function is a model dependent parameter.
Even so, valuable information can be obtained starting from basically  two observational parameters, the redshift, and 
the apparent magnitude. In this paper, we have made an independent analysis on the Hubble diagram of extragalactic high-z objects (QSOs and AGNs), on the basis of the more recent data as possible. 

It is possible to observer in the SDSS data a ``duck beak'' in the correlation between redshift and the apparent magnitude of quasars
at $z\sim 3$ as a transition for a flat correlation. Similar  upside ``duck beak'' at $z\sim 3$ is also observed in the   $Veron\& Veron$ data, in the correlation between redshift and the luminosity function obtained using the $\Omega$ model. 
In addition the high quality quasar FBQS sample, has been extended using the data of several X-ray AGNs.

In general, the result for QSOs and X-ray AGNs at high redshift ($z>3$) is  a relative flat correlation between the apparent magnitude and redshift, 
as well as between the luminosity function vs redshift,
and that is reflected in the correlation between the magnitude modulus vs redshift.
This result needs further confirmation, because the statistic of high redshift objects is still poor.
If confirmed, this result is hard to conciliate with the standard cosmological model.

\acknowledgments

This work is supported by the Brazilian National Council for Research (CNPq), under Grants No. $479813/2004-3$ and
476498/2007-4. We are grateful to various QSOs and AGNs catalogs available on the web and to their open data policy, especially to Quasars FBQS, SDSS and V$\&$V catalogs.
For additional information on this work, please write to
\email{navia@if.uff.br}.

\clearpage

\begin{figure}
\epsscale{.80}
\plotone{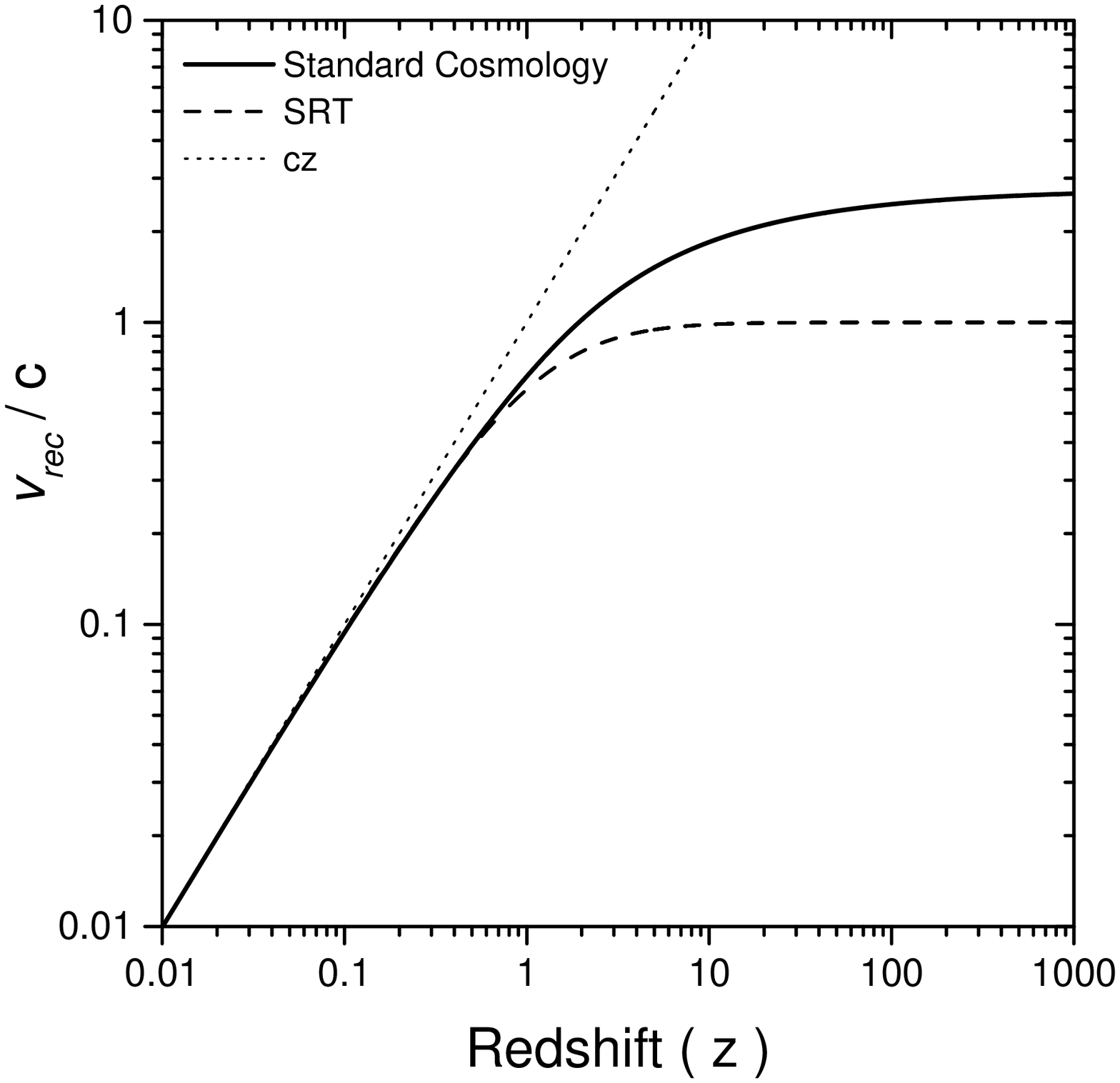}
\caption{Predictions for the recession velocity of astronomic objects  as a function of the redshift: Solid line represent the standard cosmological model prediction, under the assumption of  a flat universe. The Special Relativity Theory (SRT) is represented by dashed curve, and the dotted curve represent
the $cz$ approximation.\label{fig1}}
\end{figure}

\clearpage

\begin{figure}
\epsscale{0.80}
\plotone{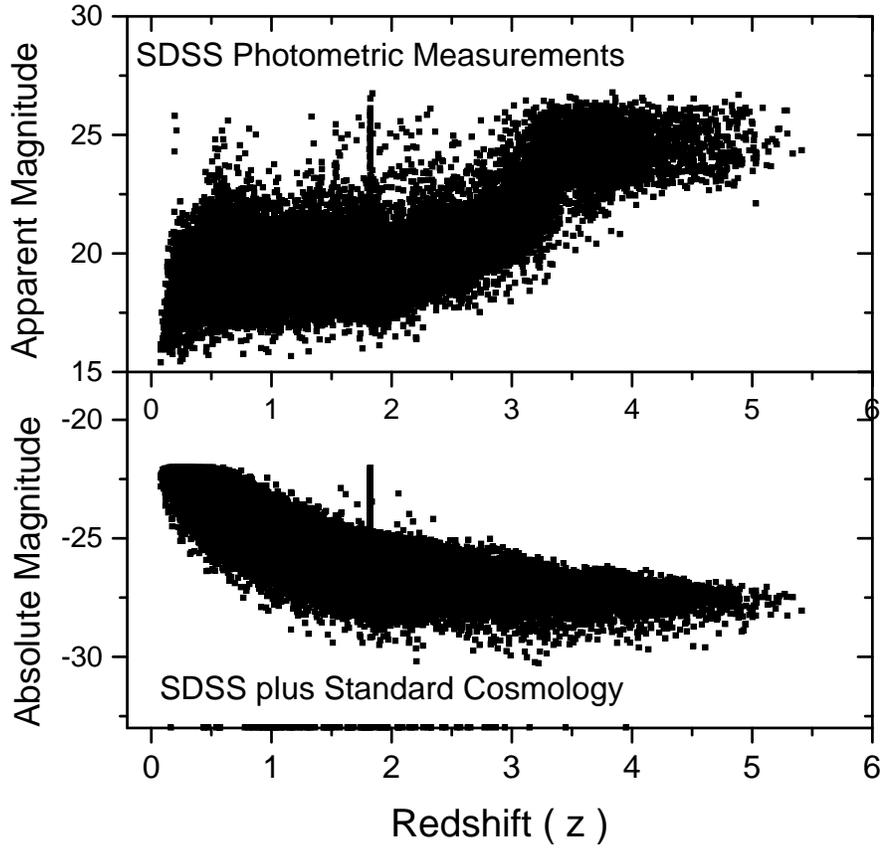}
\caption{Top panel: the apparent magnitude, the so called PSF magnitude obtained in the SDSS Release 5 photometric measurements \citep{adelman07} against the redshift. Bottom panel: the absolute magnitude as obtained by the SDSS
Release 5 photometric data 
plus the standard cosmology model ($\Omega_M=0.3$, $\Omega_{\Lambda}=0.7$ and $H_0=70km\;s^{-1}Mpc^{-1}$) against the redshift.\label{fig2}}
\end{figure}

\clearpage

\begin{figure}
\epsscale{.80}
\plotone{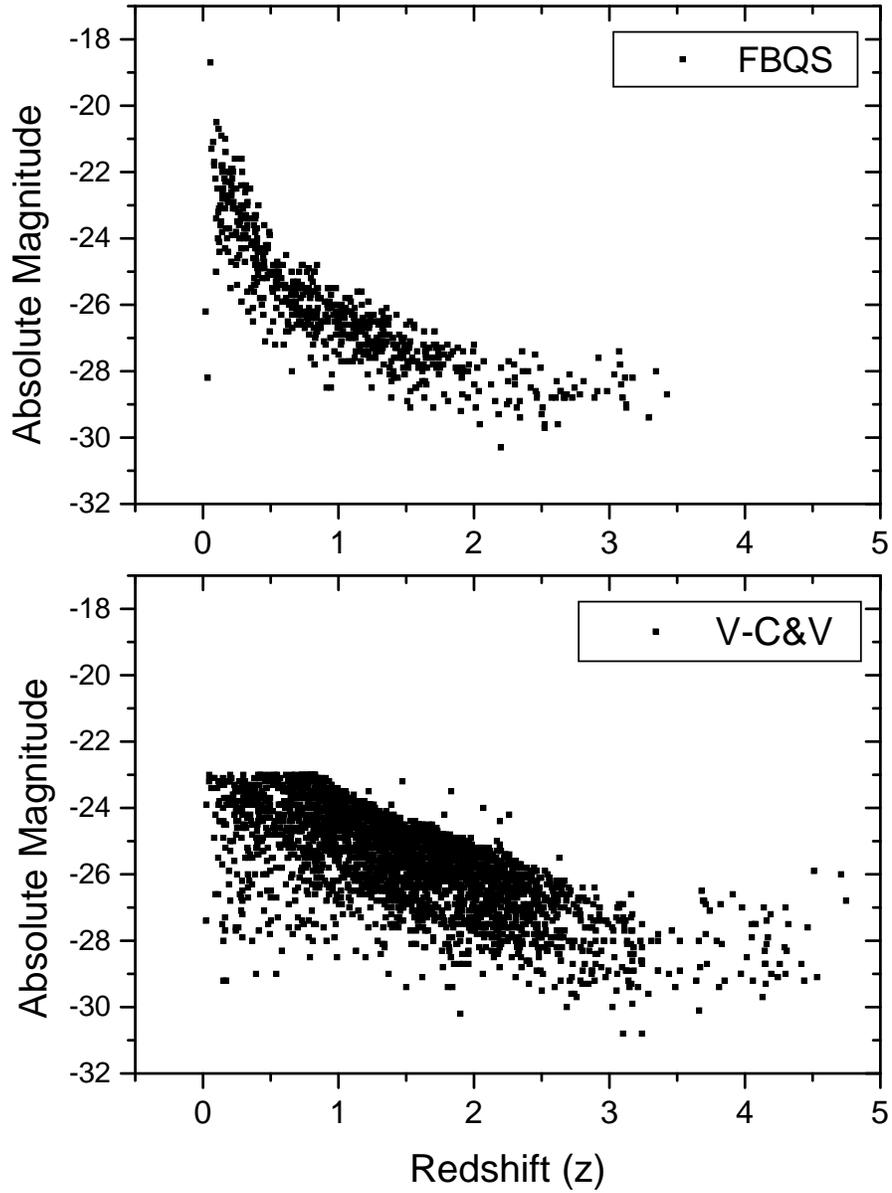}
\caption{Top panel: the absolute magnitude vs the redshift in the FBQS sample. Bottom panel: the the absolute magnitude vs the redshift in the V$\&$V sample. 
In both cases, the $\Omega$ model has been combined with the data to obtain the absolute magnitude. \label{fig3}}
\end{figure}

\clearpage

\begin{figure}
\epsscale{0.80}
\plotone{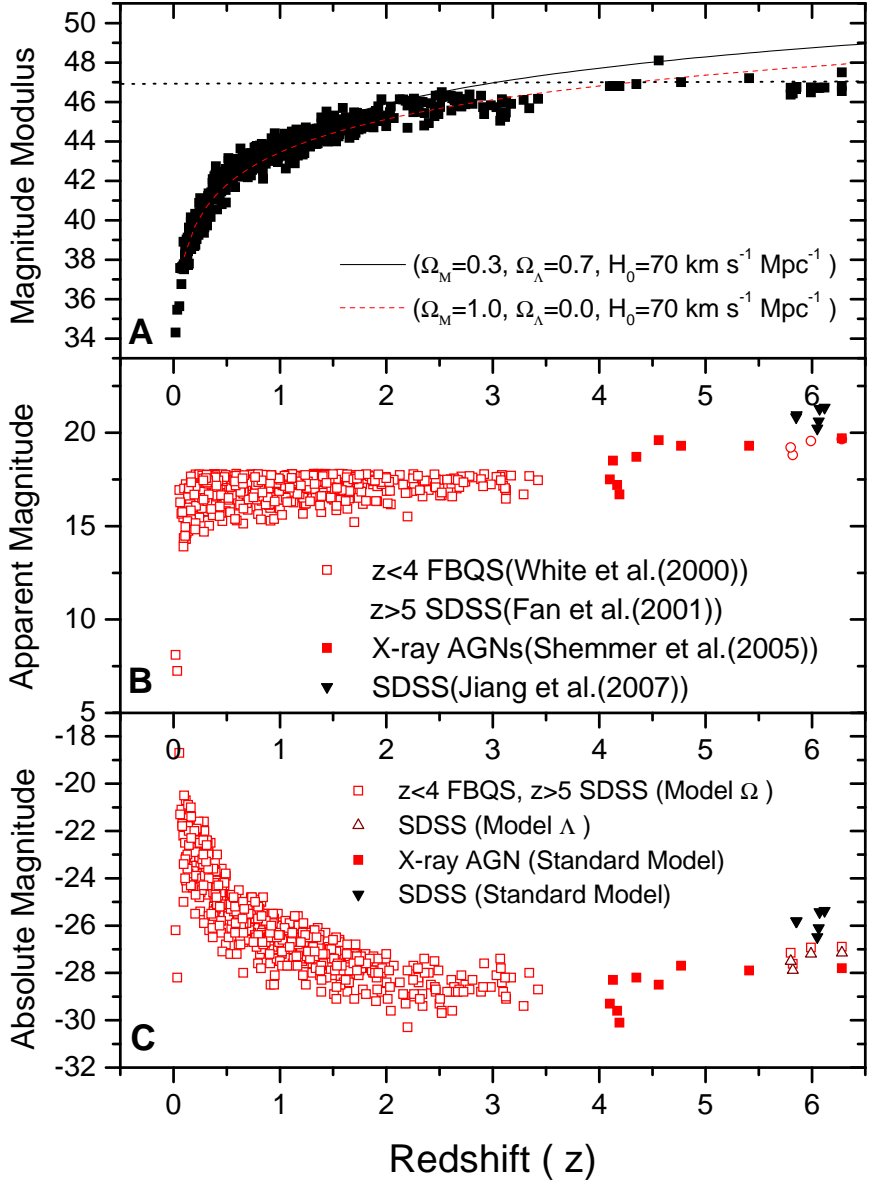}
\caption{Top panel: The magnitude modulus $m-M$ obtained combined data with cosmological models (see below). For comparison it is included the prediction of the standard cosmology model as solid-black line and the dashed-red line for two sets of cosmological parameters.
Central panel: the apparent magnitude data of several samples and in the bottom panel, their absolute magnitude for quasars and AGNs respectively. \label{fig4}}
\end{figure}

\end{document}